\documentclass[conference]{IEEEtran}
\IEEEoverridecommandlockouts
\usepackage{cite}
\usepackage{amsmath,amssymb,amsfonts}
\usepackage{algorithmic}
\usepackage{graphicx}
\usepackage{textcomp}
\usepackage{xcolor}
\def\BibTeX{{\rm B\kern-.05em{\sc i\kern-.025em b}\kern-.08em
    T\kern-.1667em\lower.7ex\hbox{E}\kern-.125emX}}
\begin{document}

\title{Pathways to Quantum: Fostering High School Student Interest in Quantum Information Science\\
\thanks{George Mason University's Building the Quantum Workforce program was developed under Grant S215K220031 from the U.S. Department of Education. However, those contents do not necessarily represent the policy of the U.S. Department of Education, and you should not assume endorsement by the Federal Government.}}

\author{\IEEEauthorblockN{Jennifer R. Simons}
\IEEEauthorblockA{\textit{George Mason University} 
\\
Fairfax, Virginia \\
jsimons2@gmu.edu\\
}
\and
\IEEEauthorblockN{Nancy Holincheck, PhD}
\IEEEauthorblockA{\textit{George Mason University} \\
Fairfax, Virginia \\
nholinch@gmu.edu \\
}
\and
\IEEEauthorblockN{Jessica L. Rosenberg, PhD}
\IEEEauthorblockA{\textit{George Mason University} \\
Fairfax, Virginia \\
jrosenb4@gmu.edu \\
}
\and
\IEEEauthorblockN{Laura M. Akesson}
\IEEEauthorblockA{\textit{George Mason University} \\
Fairfax, Virginia \\
lakesson@gmu.edu\\
}
}

\maketitle

\begin{abstract}
This study examines the Pathways to Quantum Immersion Program at George Mason University, which introduces high school students to quantum concepts and careers. Fifteen group interviews with a total of 45 students were analyzed to examine how the program supported student awareness, knowledge, and perceptions of quantum careers. Participants gained a better understanding of various quantum and quantum-adjacent careers, as well as the paths they might follow to pursue these careers. Interactions with professionals helped adjust academic and career expectations, solidifying student interest in these fields. Peers and mentors played crucial roles in students' learning journey. 

\end{abstract}


\section{Introduction}
Quantum information science (QIS) is a rapidly evolving field that has the potential to revolutionize numerous fields through technological advances in computing, sensing, and communication.  As quantum technologies have developed, there has been a corresponding increase in demand for industry professionals, researchers, and technologists with the skills needed to support the quantum industry \cite{b1}. Current educational opportunities have not met this demand, and industry professionals have identified the scarcity of talent as the biggest challenge they face \cite{b2}.

Although quantum concepts have historically been taught only in upper-level undergraduate and graduate physics courses, the past decade has seen rising interest in introducing quantum concepts to K-12 students. In the United States, most high school and college students are unaware of the range of options for quantum careers \cite{b3} \cite{b4}, which serves to limit the number of students considering them.  High school students who are introduced to quantum concepts and careers early on can gain a competitive edge in their academic and professional pursuits, as QIS is poised to create numerous high-demand careers and drive workforce development in the coming decades. However, students' lack of knowledge of quantum career options also means that they do not know what majors and experiences they should pursue to enter those careers. 

Faculty in physics and education at George Mason University collaborated with the Potomac Quantum Innovation Center to create the Pathways to Quantum Summer Immersion Program (Pathways) in order to introduce rising high school seniors to quantum concepts and careers. High school students who were selected for the Pathways program first completed a two-week virtual course focused on quantum science concepts and applications, followed by an on-campus week visiting local university, industry, and government labs. Students had the opportunity to hear from industry professionals and scientists, as well as people working in quantum policy, to support student understanding of the possible paths to a career in a quantum or a quantum-adjacent field. Following the in-person week, students were invited to extend their learning by completing an independent Quantum Vision project and presenting it at the Quantum World Congress. 

This study explores how student engagement with the Pathways to Quantum Immersion program increased student interest and awareness of QIS careers, helped students set realistic academic and career expectations, and provided further action steps for students in pursuing QIS and QIS-adjacent goals.  

\section{Conceptual Framework and Research Questions}

We used social cognitive career theory (SCCT) \cite{b5} as a framework for understanding how the Pathways program supports student thinking about quantum careers. SCCT asserts that career development is influenced by the interaction between self-efficacy (beliefs in one’s ability to succeed), outcome expectations (beliefs about the consequences of one's behavior), and students’ personal goals. To progress on a given career path, students must be interested in the field, believe that it is one in which they could find success, set related career goals, and act toward their chosen career. SCCT also emphasizes the role of contextual factors in shaping career development.

\begin{figure*}[!ht] \centering \includegraphics{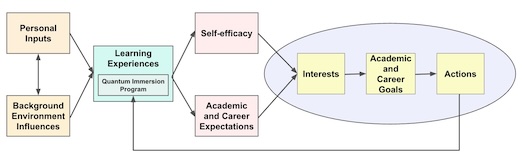} \caption{Social Cognitive Career Theory Model for Quantum Immersion Program} \label{fig:test} \end{figure*}

Learning experiences are essential in creating interest in QIS career pathways, but interest alone is insufficient. These learning experiences also need to influence students’ self-efficacy and outcome expectations. Brown and Lent \cite{b6} posit that students' beliefs about specific careers may be faulty, leading them to eliminate them as career options. Carpi et al. \cite{b7} hold that as students learn more about career paths in science they often have to re-assess their working conception of STEM careers. This is amplified in fields like QIS, where if students are aware of various quantum careers, they often view them only for the most brilliant students \cite{b4}.

The Immersion program provided a learning experience aimed at increasing student self-efficacy and redefining academic and career expectations for QIS (see Figure 1). The goal of balancing QIS concepts and career options in the Pathways program is to raise QIS awareness and interest, not to build deep student knowledge of QIS content. The program also aimed to help students understand steps they can take to progress towards a career in QIS. As a result, student interest in STEM careers, particularly QIS, should be impacted, potentially resulting in the redefinition of their future goals.  

Using social cognitive career theory, this paper focuses on how engagement with the Pathways program has increased student interest in and awareness of QIS careers, helped students set realistic academic and career expectations, and provided further action steps for students in pursuing QIS and QIS-adjacent goals. This paper addresses the following research questions:
\begin{enumerate}
 \item What awareness and knowledge of QIS and QIS-adjacent careers do students in the Pathways program have?
 \item How did participation in the Pathways program impact students’ thinking about QIS careers?
\end{enumerate}

\section{Methods}

\subsection{Program Context}

Pathways is a partnership between George Mason University, a large public R1 minority-serving institution in Northern Virginia, and the Potomac Quantum Innovation Center, founded by Connected DMV, a non-profit in the DC region. Rising high school seniors are admitted for a 3-week program completed as a cohort over the summer. Admission is competitive, requiring short answer essays and teacher recommendations. It is free for participants, and students who complete the program receive a \$500 stipend.

The program is broken into two distinct sections, followed by optional components. The first section is an asynchronous online program that introduced students to basic quantum concepts and careers over approximately two weeks. Modules included wave-particle duality, superposition, quantum states, quantum measurement, entanglement, quantum applications, and quantum careers. Each module included multimodal delivery of material including videos, simulations, readings, and formative quizzes. Our team drew on existing resources readily available online, including YouTube videos and PhET simulations. Students were required to further explore each of the module concepts individually and share an online resource with their peers through a Padlet. Minimal mathematics concepts were included in the online program to increase accessibility for students. The online component was designed to provide students with background information for the program's second component. 

The program's second component consisted of a 1-week residential course, designed to reinforce concepts learned during the virtual course while also providing students with an opportunity to learn first-hand about QIS careers. During this week, students spent time in the classroom exploring the basic concepts from the virtual week in more depth through lectures and labs. Most of the week was spent on field trips to local quantum labs or industry sites. Students engaged with experts in the field including scientists, engineers, entrepreneurs, and policymakers. Locations visited varied by year, but included quantum and quantum-adjacent locations such as laboratories at George Mason University and the University of Maryland, MITRE, NASA Goddard, and IonQ. In these locations, students got to visit facilities where quantum research was being conducted, heard from scientists about their current research projects, and their educational and career paths. Students also learned about quantum policy through visits with the White House Office of Science and Technology Policy or think-tanks like the Information Technology and Innovation Foundation.  

Optional parts of the program after the in-person week included micro-internships with local area partners in fields including quantum education, quantum materials research, quantum computing, and nanomaterials research. Students also had the opportunity to create quantum vision posters under the guidance of George Mason faculty. These posters were presented to experts at the Quantum World Congress. 

\subsection{Participants}\label{BB}
The Pathways program was advertised to schools and school districts across the Northern Virginia, Washington, D.C., and Maryland region through the professional networks of the faculty on the project and Connected DMV. The research team purposefully reached out to a broad range of schools and districts, especially those with large populations of minoritized students in STEM and Title I schools in which students may have fewer opportunities. The advertisement noted that no prior knowledge of quantum was necessary. The research team purposefully selected students who had few quantum-related opportunities through their schools but demonstrated an interest in STEM and in learning more about quantum and quantum careers. Twenty-four students were selected to participate in 2023 and again in 2024, representing nine regional school districts. Their demographics are presented in Table 1. 

\begin{table}[htbp]
\caption{Participant Demographics}
\begin{center}
\begin{tabular}{|l|c|c|}
\hline
 & 2023 Participants & 2024 Participants \\

\hline
Total Participants	&23	&21 \\
Race/Ethnicity& & \\
\hspace{1em}African American/Black &3 &4\\
\hspace{1em}Asian or Asian American &13 &9\\
\hspace{1em}Hispanic or Latino/a/e &1 &2\\
\hspace{1em}Caucasian/White &5 &5 \\
\hspace{1em}Other &1 &1 \\
\hspace{1em}Unknown &1 &2\\
Gender & & \\
\hspace{1em}Female &10 &7\\
\hspace{1em}Male &13 &11\\
\hspace{1em}Nonbinary &0 &0\\
\hspace{1em}Not reported &0 &2\\
\hline
\end{tabular}
\label{tab1}
\end{center}
\end{table}

\subsection{Data Sources}\label{CC}
Data sources included student application materials, an end-of-program survey, and focus-group interviews conducted at the end of the in-person immersion week. 

Student applications included responses to Likert questions regarding their knowledge of quantum and careers, short answer questions reflecting their interest in quantum and the Pathways program, and teacher recommendations. Previous research experience and participation in STEM activities were also collected. The end-of-program survey was completed on the final day of the in-person program that included feedback on the program and open-ended demographic questions related to race and ethnicity and gender. The program feedback was provided through Likert questions about students’ learning during the various components of the program and open-ended questions about their interest in quantum topics and quantum careers.

The research team conducted focus-group interviews at the end of the week-long in-person component. Groups consisted of 3-4 students to encourage discussion among participants. Interviews were transcribed using an automated transcription tool and checked for accuracy by research team members. Interview questions included general impressions of the virtual and in-person components of the program, student confidence in QIS concepts, and their interest in QIS and STEM careers. 

\subsection{Data Analysis}\label{DD}
Data was analyzed using qualitative coding. The researchers read through the eight focus group interviews in their entirety before open-coding. Codes developed through open-coding included thoughts about STEM careers, thoughts about non-STEM careers, confidence in STEM, confidence in academics, misconceptions of quantum, interest in quantum (sub-codes included computing, engineering, and physics), prior experience with quantum (sub-codes included education, pop culture), career goals, student descriptions of quantum careers and pathways, academic hopes/plans for future related to quantum, and social factors (sub-codes included peers and mentors). Second-cycle analysis themes were developed from the coding process \cite{b8}. 

Analysis of survey responses focused on student responses to two open-ended questions: “12. What are your college and career plans after high school?” and “13. In what ways has the Quantum Immersion Program influenced your thinking about college or your careers?” Responses were coded using structural coding to draw general themes. Researchers then engaged in a second round of iterative coding, using the codes developed during the first round \cite{b8}. 

\section{Findings}
Social cognitive career theory \cite{b5} indicates that learning experiences can develop self-efficacy and academic and career expectations. Self-efficacy and expectations can, in turn, influence or be influenced by contextual factors including interests, academic and career goals, and steps taken to pursue new learning experiences. Students indicated that the Pathways program informed their academic and career goals. Specifically, students noted (a) an increased awareness of and interest in QIS careers, (b) adjusted academic and career expectations after observing professionals, and (c) intentional action steps they planned to take to achieve their newly stated academic and career goals. The Pathways program helped inform students' choices about future career paths.

\subsection{Increased student awareness of and interest in QIS careers}\label{AA}
At the end of the Pathways program, students overwhelmingly noted an increase in awareness of potential QIS careers and interest in pursuing QIS or QIS-adjacent careers. Prior knowledge of quantum career paths was often confined to quantum computing or physics. After the Pathways experience, students could identify QIS careers outside of a quantum physicist or computer scientist. One noted, “It greatly opened my mind to the different options and applications of quantum. I am now more excited to work with others in adjacent careers and fields to see the widespread applications.” This reflected students’ experiences, as another student explained, “I had only previously heard of quantum computing and I didn't know there were any other jobs in quantum, but now I know that there's quantum materials, quantum cryptography, quantum sensing. So there's a lot of fields within quantum that you can go into for research.” Students also saw the value of QIS careers outside research with many surprised that it has a “big role in industry.” Similarly, another mentioned how exposure to careers and research “fully broadened my awareness of what it could possibly be.” 

As student awareness of the wide variety of QIS and QIS-adjacent careers increased, their interest in pursuing those careers also increased. One student mentioned that before the program they had little interest in pursuing a QIS career, but now they may minor in QIS while pursuing a quantum/aerospace career. Another admitted “This program influenced many aspects of my career. It allowed me to acknowledge a field that I was never really interested in and want me to genuinely pursue in it.” Several students indicated they would pursue additional classes or minors in QIS or QIS-adjacent fields, like material science. 

Furthermore, after interacting with professionals, students also became interested in the pathways to enter these QIS careers. Multiple students indicated an interest in master's degrees in relevant fields, with several wanting to pursue PhDs. As one student stated, “This program really opened my eyes in terms of careers because I would've never even imagined that something as big as quantum was being kept somewhat a secret from young people. It also gave me a boost in confidence to pursue higher education and get a PhD."

Overall, the Pathways program increased awareness of the various careers in STEM and QIS. Most of the students (54 of 57) retained their interest in a STEM major and career at the end of the Immersion program, with 15 students explicitly indicating an interest in continuing to advance their QIS education after completing the Pathways program. Of those 15, five students explicitly stated that they see themselves in a quantum career. 

\subsection{Adjusted Academic and Career Expectations}\label{BB}
In addition to not knowing the various QIS careers available, many students began the program without understanding what STEM careers entail. The Pathways program allowed students to benefit from real-world experience and learn about professionals' academic and career paths. This led to students developing more realistic academic and career expectations. In particular, students set high but flexible expectations for their academic paths. One student described how the experience reframed her perspective on college selection, “we talked to people who went to Yale and GMU and were still working at the same place so it’s helped me be less anxious I guess.” Students also indicated that interacting with professionals helped them understand how college could help prepare them for their careers as it helped her understand “how what I learned in college can be applied in a career sense.” Overall, many students indicated more flexibility in their academic expectations.

Engaging with the professionals also exposed students to the realities of various QIS careers, which made them appear “less science fiction.” The experience provided students with a more holistic view of the field, allowing them to see “the relationship between research, development, and industry.” The interaction with professionals reinforced some students’ interests. Still, many students found themselves readjusting their academic and career goals. One student admitted to shifting their interest away from entering academia after finding out how much time it takes and the risks involved. Another student expressed a new interest in a science career, noting “it changed my perspective seeing people who've came [sic] out of science majors and seeing them putting it into good use.” Beyond the day-to-day job-related realities, one student also noted that this experience made STEM careers less intimidating to pursue, “the STEM field always seemed really hard to enter, but I learned and actually believed that with the right network, the right education, and the right mindset I can enter and excel in the industry.” Overall, interacting with professionals and exposure to QIS careers helped change students’ expectations of not just QIS careers, but STEM careers in general.

Connecting with QIS professionals also helped students feel more comfortable adjusting their academic and career expectations to their interests. “[the Pathways program] has reassured me that the path to a good career is not always linear and I will have many opportunities to explore topics that interest me.” Many students noted that the experience both broadened their academic interests and narrowed their career interests. In discussing the integration of QIS and medicine, one student declared, “this program has narrowed down some of the careers I'm more interested in, as well as really open [sic] my eyes on how real and soon some of these careers are happening.” After entering the program interested in research, another student realized they were “much more interested in being an engineer than a researcher.” The Pathways program allowed students to engage more fully with QIS careers and how they might pursue them. Ultimately, this led to students either deepening or reevaluating their career interests. 

\subsection{Intentional Action Steps for Students in Pursuing QIS Goals}\label{CC}

Due to the nature of the Pathway program, students were able to learn more about how QIS professionals entered the field. Many students were inspired by what they heard and began to map out next steps to pursue their goals. Some students focused on short-term actions that could be taken in their senior year. One student mentioned connecting with other students and starting a Quantum Club at their high school. Several students became even more intentional in selecting universities of interest, with multiple students indicating that they were explicitly looking for universities that offer quantum classes or quantum labs. Others began researching alternative majors that may complement their initially planned major. One student noted, “I was actually originally only considering CS, but now after seeing everyone here has physics, material science engineers, a lot of engineers, so now I'm considering a double major. Most likely CS and physics.” Other students started mapping out further learning experiences. Several students felt they could now reach out for more information from experts and further hands-on learning opportunities. Students named several industry partners for future internship experiences. 

Interestingly, several students mentioned their perceived limitations in pursuing their goals and what actions they felt could help them overcome these impediments. One student found that visualization was incredibly important in overcoming their disinterest in math, 
\begin{quote}
“But seeing a lot of these people who also said that they didn't like math, but they learned to love it… So I was like, okay, even if I'm bad right now, I can learn to be good. I guess I saw myself in, like, five years, okay, I can also be like this person if I work hard enough.”
\end{quote}
Other students identified the importance of physics or coding and expressed their desire to complete additional coursework, even if they disliked the specific content area, like the student who declared, “I don't like physics. I hate physics,” who later stated, “in college I'll likely take more physics courses, take more math courses, see what I'm interested in, try to branch out into STEM.” This sentiment was particularly strong across the six students who acknowledged an interest in pursuing a doctorate. Of those six, four identified content areas they found useful but uninteresting.

\subsection{Importance of Peer and Mentor Relationships}\label{DD}

Students frequently cited the importance of both peers and professionals in shaping and reshaping their interest in QIS. Several students noted the importance of peers in their learning journey, with one reflecting, “Working with people our age and breaking down information together was invaluable. We've formed many meaningful bonds and friendships through this, which I consider to be the most important aspect of the program.” Students also found networking with professionals important in making QIS careers more realistic and attainable. 

\section{Discussion}
This paper focused on how the Pathways program impacted high school participants’ perceptions of QIS and QIS-adjacent careers. We found that the Pathways program raises awareness of QIS and QIS careers among the high school participants. Their knowledge of different QIS and QIS-adjacent careers increased considerably, from generalized concepts of quantum physicists to professions including engineering and policy-making. Using SCCT to understand student inclinations toward QIS further, we noted that the program helped students adjust their academic and career expectations through meaningful engagement with professionals.

Multiple students agreed that the Pathways program gave students a much better understanding of specific careers that are often shrouded in secrecy or glamorized by popular culture. Adjustment of career expectations further refined student interest in QIS, resulting in adjusted academic and career goals. With these adapted goals, students could better plan next steps to advance toward a STEM career. Overall, student interest in QIS was piqued by this project, with 15 of the study’s 45 students indicating an interest in a QIS-related career. They indicated various next steps, including the development of a club, a change in intended major, or planning another learning experience (e.g., internships). 

Participating in the Pathways program impacted students’ academic and career expectations, interest in QIS, and planning for future actions to accomplish academic and career-related goals. Overall, we found this program significantly impacted students’ interest in pursuing QIS careers. We acknowledge that the information analyzed only reflects student interest at the end of an immersive experience. In the upcoming years, we plan to follow up with program participants to assess the longer-term impact of the Immersion program and determine if participant interest in QIS is sustained over multiple years. 

\section{Conclusions and Implications}

The findings from this study have important implications for educational practices aimed at increasing awareness and interest in Quantum Information Science (QIS) careers among high school students. The Pathways program demonstrated that immersive experiences, such as interactions with professionals in the field and exposure to diverse QIS-related careers, can significantly impact students' academic and career expectations. These findings suggest that educators should prioritize providing opportunities for authentic engagement with experts and career exploration in QIS to help students develop clearer and more informed career goals. Furthermore, incorporating QIS into STEM curricula at the high school level could help bridge the knowledge gap and spark student interest in this emerging field.

This study contributes to the growing body of research on effective strategies for raising awareness and interest in STEM fields among underrepresented populations, particularly those related to QIS careers. Future studies may investigate the role of peer networks and mentorship programs in sustaining student interest over time, as well as the impact of long-term follow-up initiatives on students' academic and career outcomes. Additionally, further research could examine the factors that influence students from underrepresented backgrounds to pursue careers in QIS and identify interventions designed to address potential barriers to entry. Overall, this research highlights the importance of creating immersive learning experiences for high school students that foster understanding and genuine interest in QIS careers.

\section*{Acknowledgment}
\small{George Mason University's Pathways to Quantum Immersion program benefits from the support from Connected DMV's 
George Thomas and Kieran Collinson and George Mason University's Melinda Ryan, Dr. Ben Dreyfus, and Dr. Patrick Vora. We also thank the many industry, academic, and government professionals who shared their knowledge and experiences with Pathways students, and the GMU PhD in Education students who supported data collection.}

\end{document}